\documentclass[superscriptaddress,nobibnotes,amsmath,amssymb,notitlepage,twocolumn,prl]{revtex4-2}

\usepackage{dirtytalk}
\usepackage{mathptmx}
\usepackage{graphicx}
\usepackage[caption=false]{subfig}
\usepackage{hyperref}
\hypersetup{colorlinks=true, linkcolor=blue, citecolor=blue, urlcolor=blue}

\graphicspath{{./img/}}

\newcommand{\beq}[1]{\begin{equation}\label{#1}}
\newcommand{\eep}{\;.\end{equation}}
\newcommand{\eec}{\;,\end{equation}}
\newcommand{\eeq}{\end{equation}}

\newcommand*\chem[1]{\ensuremath{\mathrm{#1}}} 

\newcommand{\G}{\Gamma}

\DeclareMathAlphabet{\mathcal}{OMS}{cmsy}{m}{n} 
\newcommand{\C}{\mathcal{C}}   

\newcommand{\MBT}{\chem{MnBi_2Te_4}}
\newcommand{\MBST}{\chem{MnBi_2S_2Te_2}}
\newcommand{\VWS}{\chem{V_2WS_4}}

\newcommand{\epc}{\epsilon_{\rm c}} 

\newcommand{\muB}{\mu_{\rm B}} 

\begin{document}

\title{Switching topological states via uniaxial strain in 2D materials}


\newcommand{\HarvardPhysics}{Department of Physics, Harvard University, Cambridge, Massachusetts 02138, USA}
\newcommand{\HarvardSeas}{John A.~Paulson School of Engineering and Applied Sciences, Harvard University, Cambridge, Massachusetts 02138, USA}
\newcommand{\MIT}{Department of Physics, Massachusetts Institute of Technology, Cambridge, MA, USA}

\author{Joshua J. Sanchez}
\email{sanchezx@mit.edu}
\affiliation{\MIT}

\author{Raagya Arora}
\email{raagya@g.harvard.edu}
\affiliation{\HarvardSeas}

\author{Daniel Bennett}
\affiliation{\HarvardSeas}

\author{Daniel T. Larson}
\affiliation{\HarvardPhysics}

\author{Efthimios Kaxiras}
\affiliation{\HarvardSeas}
\affiliation{\HarvardPhysics}

\author{Riccardo Comin}
\affiliation{\MIT}

\begin{abstract}
In topological materials, dissipationless edge currents are protected against local defect scattering by the bulk inverted band structure and band gap. 
We propose that large uniaxial strain can effectively switch a 2D Chern insulator to a topologically trivial state. 
Further, we suggest that the boundary between strained and unstrained regions of a sample can act as a new edge for dissipationless current flow. Using density functional theory (DFT) calculations we demonstrate the strain-tunability of the monolayer \MBST~band structure and the switching of the Chern number. We combine uniaxial and biaxial strain results to map out the strain-tuned topological phase diagram. 
\end{abstract}

\maketitle

\section{Introduction}
In a topological insulator, electronic bands of distinct symmetry cross and hybridize, which opens an inverted bulk band gap. 
A consequence of this mixed-symmetry band gap is the formation of chiral spin-polarized metallic bands localized at the boundaries of the inverted region, which results in conductive edge modes. 
The bulk band gap protects these edge modes against scattering, allowing a current to flow with zero dissipation. 
This topological protection of the edge mode conductivity is a hallmark of these materials, and has generated great interest for potential energy-efficient device applications. 
However, it is inherently difficult to externally tune the edge current as it requires changing the topological state of the entire inverted volume, potentially limiting its utility in practical devices. In this work we use DFT to show how mechanical strain can tune the edge mode conduction of a specific subclass of topological insulators, namely those with intrinsic ferromagnetic order (often called quantum anomalous Hall insulators or Chern insulators), though our results are applicable to other systems such as nonmagnetic topological insulators and topological crystalline insulators. 

In a topological insulator, the spin-momentum locking of the edge bands result in spin-up and spin-down electrons propagating in opposite directions around the boundary of a material. The propagation direction for a particular spin species is determined by the slope of the edge band. When both spin edge bands are equally occupied, no net charge current flows along the boundary. 

A Chern insulator is defined by the Chern number $C$, which is the difference between the number of right and left-propagating edge modes at the Fermi level. A nonzero value of $C$ arises from the magnetization of the material, which pushes edge bands for spins anti-parallel to the magnetization to higher energy and leaves only parallel spin edge bands populated at the Fermi level.
This results in a net charge current along the boundary, with the edge conductivity given by the transverse conductivity equation $\sigma_{xy} = \frac{e^2}{h}C$. Therefore, the transverse conductivity can be modified by flipping the magnetization direction (which flips the sign of the Chern number) and by changing the net number of the edge bands (which changes the integer value of the Chern number). 

Topologically protected edge modes must exist at the boundary between any two regions of space with different values of $C$ \cite{vanderbilt2018}. 
The simplest such boundary is between the physical edge of a sample and the vacuum ($C=0$). 
This was first demonstrated experimentally in magnetically doped \chem{(Bi_{1-x}Sb_x)_2Te_3} thin films \cite{chang2013experimental,checkelsky2014,chang2015high}, and later in few-layer samples of \MBT~\cite{deng2020quantum,liu2020robust,ge2020high, ning2020subtle, bhattarai2024exploring}. 
A second type of boundary exists along the domain walls between regions of net  spin-up ($C>0$) and spin-down ($C<0$) magnetization within a single sample. 
Control over the direction of edge currents was demonstrated by using an external magnetic field to ``write'' areas of a sample into spin-up or spin-down states \cite{yasuda2017quantized}. 
A third type of boundary is found between areas of different thickness within a single sample. For instance, in \MBT, the boundary between a 4-layer ($C=1$) and a 6-layer ($C=2$) region was shown to support one edge mode \cite{ovchinnikov2022}.  
These preceding examples demonstrate the diversity of approaches to generating topological boundaries in Chern insulators.

While the previous examples demonstrate boundaries defined by discrete quantities (magnetization direction, layer number), a boundary could also be formed by a continuous tuning parameter. For instance, consider a  material with a doping gradient such that the heavily doped side hosted ferromagnetism ($C=1$) while the lightly doped side did not ($C=0$). Then, there would be a path somewhere in the middle of the sample along which the inverted band gap must close and a conductive edge mode must form. 
The existence of this closed band gap path is a strict requirement of the topological nature of the inverted band structure \cite{vanderbilt2018}. 

\begin{figure*}
\centering
\includegraphics[width=1\linewidth]{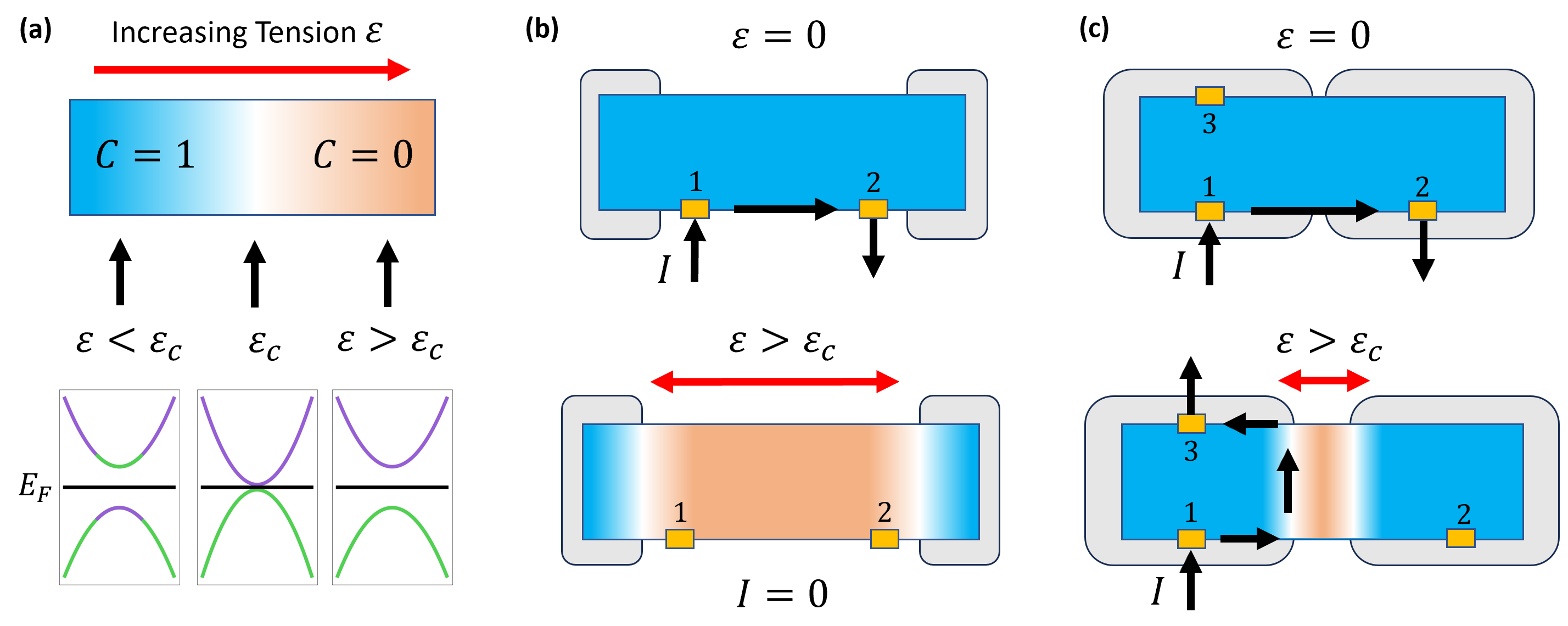}
\caption{%
{\bf (a)} Conceptual schematic of a Chern insulator with a strain-tunable band gap under a large strain gradient. 
Regions of low-strain inverted band gap ($C=1$, blue) and high-strain topologically trivial band gap ($C=0$, orange) are separated by a narrow width at the critical strain $\epc$ where the band gap is closed (white). 
{\bf (b,c)} Strain-switchable topological circuit elements. 
The sample bar is clamped to substrate plates (gray) which can move and apply strain to the sample. 
Black arrows indicate dissipationless edge current flow.
{\bf (b)} Topological transistor. 
Under zero strain, edge current flows from contact 1 to contact 2. 
Under large applied tension, the device is reduced to a trivial insulator and no current flows.    
{\bf (c)} Topological current switch. 
Under zero strain, current flows from contact 1 to contact 2. 
Under large applied strain, the current follows the internal boundary between strained and unstrained regions and is received at contact 3. 
}
\label{Fig1}
\end{figure*}

Here we introduce another type of topological boundary, which can form between regions of a sample experiencing different amounts of mechanical strain. We posit that in some Chern insulator systems, uniaxial strain may be able to close and reopen a band gap, and in doing so change the value of the Chern number. 
In that case, the boundary between regions of different strain will host conductive edge modes. 
We first discuss this process schematically for an idealized system with the desired topological strain tunability. 
Then, we discuss DFT calculations for several 2D Chern insulating materials. We show that a realistic value of uniaxial strain can indeed reverse the band gap in \MBST, suggesting that control of edge modes across a strain boundary is feasible experimentally.

We consider a $C=1$ Chern insulator with a strain-tunable band gap, where increasing tensile strain causes the topological band gap to close and reopen as a trivial gap as shown in Fig.~\ref{Fig1} (a). 
In a sample experiencing a large strain gradient, the low strain region will maintain an inverted band gap ($C=1$), while the high strain region will have a trivial band gap ($C=0$). The critical strain $\epc$ where the band gap closes marks the boundary between regions of different Chern number. At this boundary a conductive edge mode must form. 

Such a strain-tunable material could be used in a device where applied uniaxial strain controls the flow of dissipationless current. 
We discuss two strain-tunable topological circuit elements, diagrammed in Figs.~\ref{Fig1} (b) and (c). 
A topological transistor is formed by straining a large region of a sample, such that the total current flow between two contacts can be turned on and off by straining the sample between the $C=1$ and $C=0$ states. 
A topological current switch is formed by straining a narrow width between two contacts on a sample, controlling whether the unidirectional current flows around the whole sample or through the middle of the sample.

\begin{figure}[t!]
\centering
\includegraphics[width=1.0\linewidth]{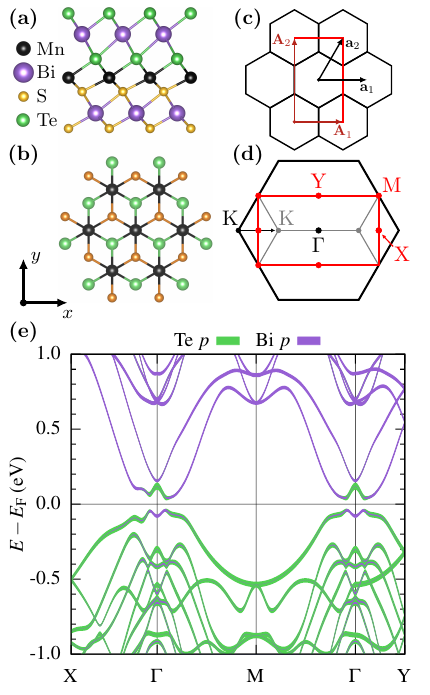}
\caption{%
{\bf (a)} Crystal structure of monolayer \MBST. 
{\bf (b)} In-plane crystal structure showing the Mn layer and its surrounding S and Te layers. 
Note the inequivalence of the $x$ and $y$ directions. {\bf (c)} Real space and {\bf (d)} reciprocal space diagrams of the hexagonal conventional cell (black) and the rectangular supercell (red) used for strain calculations. The reflected K point is shown in light gray. 
{\bf (e)} Orbital-projected band structure of \MBST~under zero strain.
The Te and Bi $p$ orbital projections are shown in green and purple, respectively.
}
\label{Fig2}
\end{figure}

\section{Results}
First-principles calculations were performed to investigate the possibility of inducing a topological phase transition with uniaxial strain (see Methods section for details).
We first explored the effect of tunable strain on \MBT~in 1 and 3 layer samples (the latter in both ferromagnetic and antiferromagnetic configurations), with initial magnetic moments of $5 \muB$ onto the Mn atoms in each layer. 
We were unable to identify a strain-induced switching between a Chern insulating and a trivial insulating state in these analyses; instead, a metalization transition usually occurred. We also investigated uniaxial strain-tuning of \VWS, a quantum anomalous Hall insulator candidate material \cite{V2WS4}, and again found a metalization transition. These results are presented in the Supplementary Materials (SM).

We then performed calculations on a monolayer of \MBST, which has the same structure as \MBT~but with half of the Te atoms replaced with S (Fig.~\ref{Fig2} (a)). Previous first-principles studies have predicted that this material hosts a doubly inverted band gap \cite{li2022interplay}.
Under biaxial tension, \MBST~was shown to undergo two topological transitions, from $C=2$ to $C=-1$, and then to $C=0$ (note that the change in sign of $C$ comes from a change in slope of the edge band with strain tuning and not from a reversal of the magnetization). In this work, we were able to identify a uniaxial strain-mediated transition from the $C=2$ phase to $C=1$, $C=0$, and metallic phases, discussed below. 

\MBST~has a hexagonal crystal lattice with $\C_3$ rotational symmetry, meaning the crystal structure is distinct along the perpendicular $x$ and $y$ directions. Here we take $x$ to be the ``zig zag'' direction and $y$ to be the ``armchair'' direction, and explore strain tuning along both. In order to model uniaxial strain, a rectangular $1 \times \sqrt{3}$ supercell was constructed and superimposed over the hexagonal lattice, see Fig.~\ref{Fig2} (c). The lattice was stretched or compressed in one direction, and the orthogonal lattice vector was scaled by a Poisson ratio of -0.3, a typical value for 2D materials.

The rectangular supercell results in zone folding of the hexagonal Brillouin zone (BZ) in momentum space (Fig.~\ref{Fig2} (d)).
The K and K' points at the corners of the hexagonal BZ map to identical points within the rectangular BZ between the X point and the $\Gamma$ point.
The M point at the midpoint of the edge of the hexagonal BZ becomes the corner of the rectangular BZ.

We first calculated the electronic structure of \MBST~under zero strain (Fig.~\ref{Fig2} (e)), and our results are in agreement with the previous report \cite{li2022interplay}. At the $\Gamma$ point, the valence band is primarily of Te $p_{xy}$ character, while the conduction band is primarily of Bi $p_z$ character. The band gap itself shows two points of band inversion, which results in an expected Chern number of $C=2$ (discussed below). 

We then calculated the electronic structure as a function of tensile and compressive uniaxial strain up to 8\% along the $x$ axis and up to 10\% along the $y$ axis; other 2D materials such as graphene and transition metal dichalcogenides have previously been demonstrated to withstand such large strain values \cite{lee2008measurement,roldan2015strain}.

\begin{figure*}[t!]
\centering
\includegraphics[width=\linewidth]{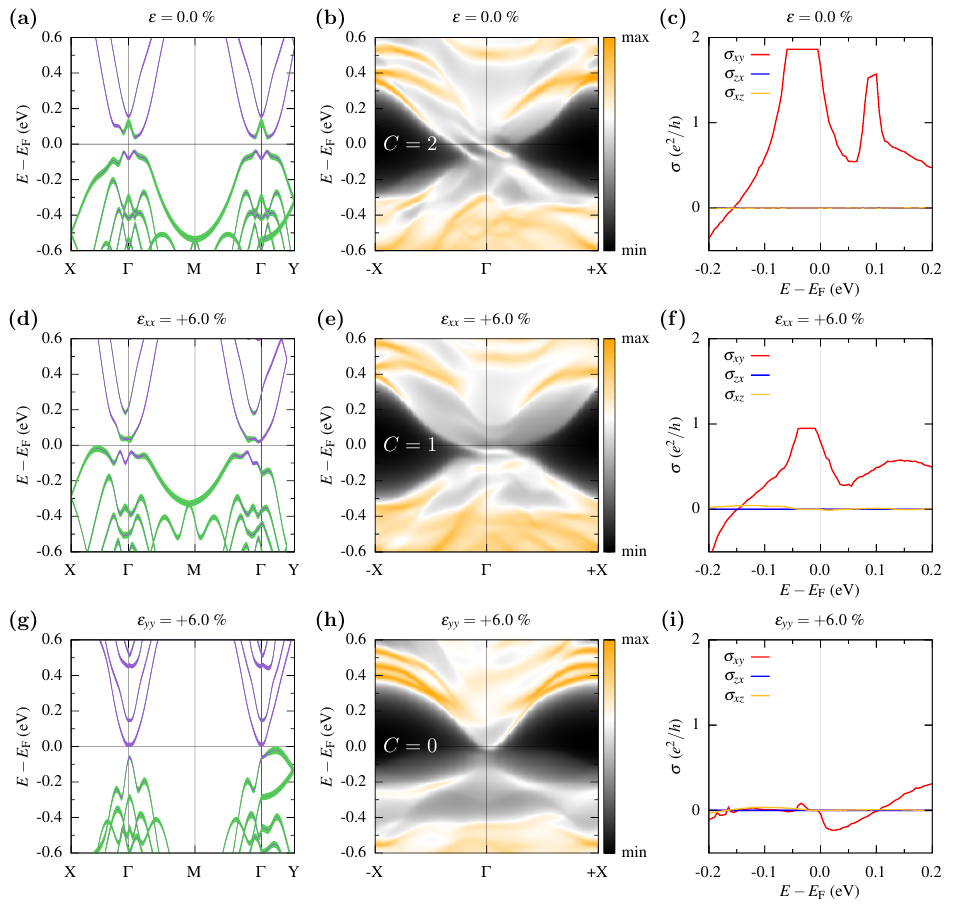}
\caption{%
Comparison of \MBST~under zero strain (a,b,c), $\epsilon_{xx} = 6\%$ tensile strain (d,e,f), and $\epsilon_{yy} = 6\%$ tensile strain (g,h,i). (a,d,g) Band
structure calculations. (b,e,h) Calculated edge states. The number of Fermi level-crossing edge modes (and corresponding Chern number $C$)
changes from two ($C = 2$) under zero strain, to one ($C = 1$) for $\epsilon_{xx} = 6\%$, to zero ($C = 0$) for $\epsilon_{yy} = 6\%$. (c,f,i) Quantum anomalous Hall
conductivity calculations. Plateaus for the in-plane transverse conductivity $\sigma_{xy}$ occur near the expected quantized conductivity values for the
$C = 2$ and $C = 1$ cases, and is nearly zero for the $C = 0$ case.
}
\label{Fig3}
\end{figure*}

We first compare results for the zero- and tensile-strain cases (Fig.~\ref{Fig3}). Under tensile strain, we observed that the band gap closes and reopens at around 3\% strain along the $x$ direction and at 3.5\% strain along the $y$ direction. This results in a band gap with only one point of inversion along $x$ (Fig.~\ref{Fig3} (d)) and topologically trivial band gap for $y$ (Fig.~\ref{Fig3} (g)). We then calculated the edge electronic band structure.  The presence of one or two edge modes in the fundamental band gap indicates the QAH phase, while the gapped edge spectrum indicates topologically trivial edge state. Under zero strain, we found two right-moving edge states that cross the Fermi level, indicating a Chern number of $C=2$ (Fig.~\ref{Fig3} (b)). For tensile strain along $x$ this was reduced to a single edge state ($C=1$) (Fig.~\ref{Fig3} (e)) while for tensile strain along $y$ the edge states vanish ($C=0$) (Fig.~\ref{Fig3} (h)). Finally, we calculated the transverse conductivity (Figs.~\ref{Fig3} (c), (f), (i)), and find peaks (or no peak) in the in-plane transverse conductivity $\sigma_{xy}$ corresponding to the expected quantized conductivity equation $\sigma_{xy} = \frac{e^2}{h}C$.

\begin{figure*}[t!]
\centering
\includegraphics[width=\linewidth]{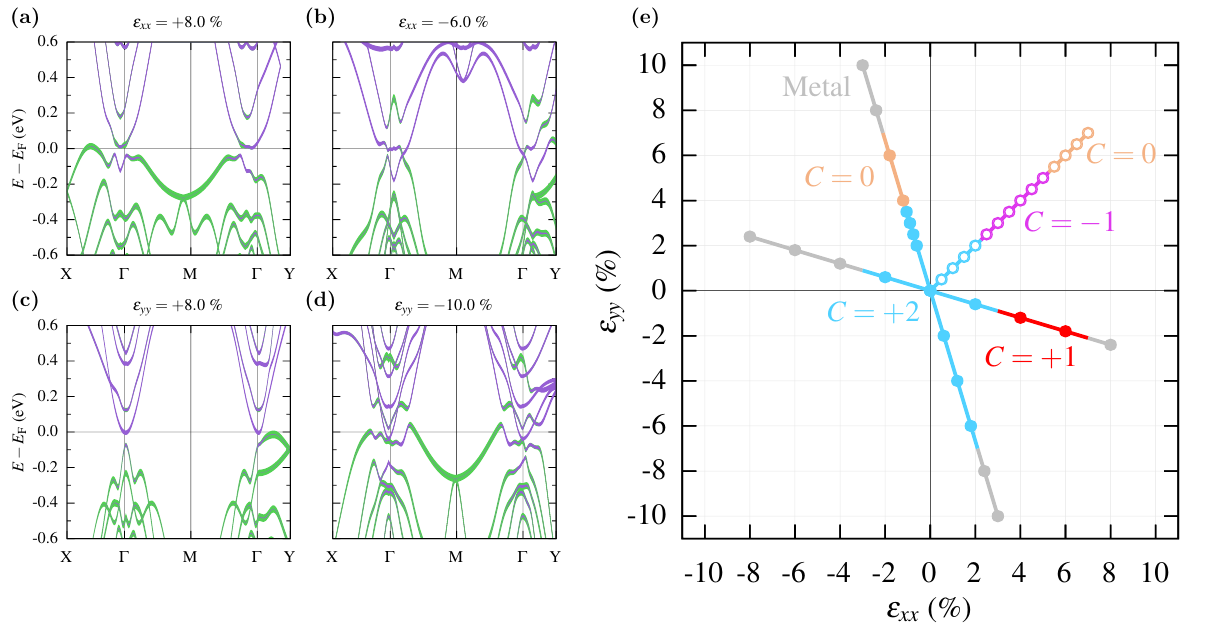}
\caption{%
{\bf (a-d)} Orbital-projected band structures under large tensile and compressive uniaxial strain along {\bf (a,b)} $x$ and {\bf (c,d)} $y$.
{\bf (e)} Topological phase diagram of \MBST~as a function of strain in the $x$ and $y$ directions.
The $C=+2, +1, 0,$ and $-1$ phases are shown in blue, red, orange, and magenta, respectively, and the metallic phase is shown in gray.
The filled circles represent calculations from this work, and the empty circles represent calculations from Ref.~\cite{li2022interplay}.
}
\label{Fig4}
\end{figure*}

In Figure 4a-d, we present DFT results at very large strain values for both tension and compression along $x$ and $y$. In all cases, a metalization transition occurs when a side band rises in energy to eventually cross the Fermi level. Then, the total conduction includes a non-dissipative source from edge conduction, and a dissipative source from regular metallic bulk conduction. Combining our results for uniaxial strain with previous results for biaxial strain \cite{li2022interplay}, we can construct a partially-explored strain-tuned phase diagram for \MBST~(Fig.~\ref{Fig4} (e)). These results show clearly that the edge conduction of \MBST~is strongly strain-tunable, and transitions from the $C=2$ state to any of $C=1$, $C=0$, $C=-1$, or a metallic phase can be obtained by choice of strain axis and symmetry.


\section{Discussion}
In \MBT, strong SOC induces band inversion and nontrivial topology in multilayers \cite{li2024progress}.
Switching on SOC results in a very small gap about the Fermi level, with significant band inversion between the Bi and Te $p$ states.
After exploring the possibility of inducing a topological phase transition in pure \MBT~\cite{deng2020quantum,liu2020robust,ge2020high}, we found that the band inversion could not be undone through the application of biaxial or uniaxial strain.
The gap and band inversion was found to be robust against biaxial strain.
Upon application of uniaxial strain, the gap closed (insulator--metal transition), but did not reopen again as a trivial insulator.
One possible reason for the persistence of the topological state (avoided crossing), is the combination of symmetries here.
Biaxial strain preserves both inversion symmetry and $\C_3$ symmetry, which may be protecting the band inversion and avoided crossing.
Uniaxial strain breaks rotation symmetry but not inversion symmetry, which may be enough to remove the avoided crossing, but not enough to induce a topological phase transition (we have a topological $\to$ metal transition).
This is in fact a common occurrence in attempts to realize topological insulators \cite{inoue2019band}. 

In \MBST, the inequivalence of the Bi-Te and Bi-S layers surrounding the Mn layer classifies it as a Janus material. This naturally breaks inversion symmetry, which may be the necessary condition that enables external structural tuning to cleanly uninvert the band gap, as seen for both biaxial and uniaxial strain. Janus materials have been gaining interest for their additional complexity over centrosymmetric 2D materials \cite{defo2016strain,zhang2022janus}. Like other Janus materials, \MBST~likely exhibits a c-axis spontaneous electrical polarization \cite{zhang2020enhancement} due to the different electronegativities of the Bi-Te and Bi-S layers. It remains unexplored how this may interact with the topological features of the system. The inequivalence of the two surfaces also is expected to result in a difference in the interlayer coupling to a substrate \cite{zhang2020enhancement,zhang2021spectroscopic}, which could affect the degree of strain transmission to the sample. While an attempt to grow crystals of \MBST~does not appear to have been reported in the literature thus far, there have been several successful attempts at growing Janus transition metal dichalcogenides \cite{zhang2022janus}, and several growth methods could be pursued. 

In conclusion, we have demonstrated the feasibility of using realistic values of applied uniaxial strain to switch off the spontaneous edge conduction in a quantum anomalous Hall material. 
This provides a path towards creating electronic devices with switchable topological states. 
Our result adds to the growing body of work exploring the structural tunability of topological materials \cite{li2022interplay,zhang2023strain,mutch2019evidence}.


~~~~~~~~~~~~~~~~~~~~~~~~~~~~~~~~`




\section{Acknowledgements}
R.A., D.B., D.T.L.~and E.K.~acknowledge the US Army Research Office (ARO) MURI project under grant No.~W911NF-21-0147 and from the Simons Foundation award No.~896626. J.J.S. acknowledges support from the National Science Foundation MPS-Ascend Postdoctoral Research Fellowship, Award No. 2138167. Any opinions, findings, and conclusions or recommendations expressed in this material are those of the author(s) and do not necessarily reflect the views of the National Science Foundation.


\section{Methods}


Density functional theory calculations were performed using the {\sc VASP} code\cite{kresse1996efficiency}, PBE exchange-correlation functional~\cite{perdew1996generalized}, and projector augmented wave (PAW) pseudopotentials \cite{blochl1994projector} with 13, 15, and 6 valence electrons for the Mn, Bi, and Te/S atoms, respectively. 
For all three materials studied we employed a plane-wave energy cutoff of 520 eV as well as the zero-damping DFT-D3  van der Waals correction~\cite{grimme2010consistent}.
Forces on each atom were reduced below 0.01 eV/\AA{} during structural relaxations, and we used a $\G$-centered 15$\times$15$\times$1 k-point grid for the final self-consistent calculations. The spin–orbit interactions are considered in the calculations.

The maximally localized Wannier transformations were performed using the Wannier90 code~\cite{mostofi2008wannier90}, with initial projections onto the $p$ orbitals of the Bi and Te atoms.
These maximally localized Wannier functions (MLWFs) were then used to build a tight-binding model using the {\sc WannierTools} code \cite{wu2018wanniertools}.
The surface states were obtained by performing a slab calculation with a $7\times 7$ supercell, using a tight binding model with the MLWFs as a basis.
The anomalous Hall conductivity was calculated using Wannier interpolation.




\bibliographystyle{apsrev4-2}
\bibliography{./references.bib}

\end{document}